\documentclass[aps,prd,onecolumn,groupedaddress,showpacs,nofootinbib,amssymb
]{revtex4}
\usepackage[dvips]{graphicx}
\usepackage{amssymb}
\usepackage{amsmath}
\usepackage{graphicx}
\usepackage{amsfonts}
\usepackage{bm}
\usepackage{color}

\providecommand{\U}[1]{\protect\rule{.1in}{.1in}}
\newcommand{\be}{\begin{equation}}
\newcommand{\ee}{\end{equation}}

\newcommand{\mincir}{\raise
-3.truept\hbox{\rlap{\hbox{$\sim$}}\raise4.truept\hbox{$<$}\ }}
\newcommand{\magcir}{\raise
-3.truept\hbox{\rlap{\hbox{$\sim$}}\raise4.truept\hbox{$>$}\ }}

\begin{document}

\title{Kinetic Scalar Curvature Extended $f(R)$ Gravity}
\author{S. V. Chervon}\email[]{chervon.sergey@gmail.com}
\affiliation{Laboratory of Gravitation, Cosmology, Astrophysics, Ulyanovsk State Pedagogical University, 100-years V.I. Lenin's Birthday square, Ulyanovsk 432071, Russia\\
Astrophysics and Cosmology Research Unit, School of Mathematics, Statistics and Computer Science,
University of KwaZulu--Natal, Private Bag X54001, Durban 4000, South Africa}
\author{A. V. Nikolaev\footnote{orcid 0000-0003-2730-8470}}\email[]{ilc@xhns.org}
\affiliation{Laboratory of Gravitation, Cosmology, Astrophysics, Ulyanovsk State Pedagogical University, 100-years V.I. Lenin's Birthday square, Ulyanovsk 432071, Russia\\
Astrophysics and Cosmology Research Unit, School of Mathematics, Statistics and Computer Science,
University of KwaZulu--Natal, Private Bag X54001, Durban 4000, South Africa}
\author{T. I. Mayorova}\email[]{majorova.tatyana@mail.ru}
\affiliation{Laboratory of Gravitation, Cosmology, Astrophysics, Ulyanovsk State Pedagogical University, 100-years V.I. Lenin's Birthday square, Ulyanovsk 432071, Russia}
\author{S. D. Odintsov}\email[]{odintsov@ieec.uab.es}
\affiliation{ICREA, Passeig Luis Companys, 23, 08010 Barcelona, Spain\\
Institute of Space Sciences (ICE,CSIC) C. Can Magrans s/n,
08193 Barcelona, Spain\\
Institute of Space Sciences of Catalonia (IEEC),
Barcelona, Spain}
\author{V.K. Oikonomou}\email[]{v.k.oikonomou1979@gmail.com}
\affiliation{Department of Physics, Aristotle University of Thessaloniki, Thessaloniki 54124, Greece\\
 Laboratory for Theoretical Cosmology, Tomsk State University
of Control Systems
and Radioelectronics (TUSUR), 634050 Tomsk, Russia\\
Tomsk State Pedagogical University, 634061 Tomsk, Russia}

\date{\today}

\begin{abstract}
In this work we study a modified version of vacuum $f(R)$ gravity
with a kinetic term which consists of the first derivatives of the
Ricci scalar. We develop the general formalism of this kinetic
Ricci modified $f(R)$ gravity and we emphasize on cosmological
applications for a spatially flat cosmological background. By
using the formalism of this theory, we investigate how it is
possible to realize various cosmological scenarios. Also we
demonstrate that this theoretical framework can be treated as a
reconstruction method, in the context of which it is possible to
realize various exotic cosmologies for ordinary Einstein-Hilbert
action. Finally, we derive the scalar-tensor counterpart theory of
this kinetic Ricci modified $f(R)$ gravity, and we show the
mathematical equivalence of the two theories.
\end{abstract}

\pacs{04.20.Jb, 04.50.Kd}
\keywords{$f(R, (\nabla R)^2) $ gravity, cosmological solution, scalar-tensor equivalence}

\maketitle


\section{Introduction}

Modified gravity in it's most general forms, serves as a formal
theoretical framework which can potentially harbor in a consistent
way both cosmological and astrophysical phenomena
\cite{reviews1,reviews2,reviews4,reviews5,reviews5a,reviews5b,reviews6,reviews7}.
For example, in the context of the most sound and simple modified
gravity, $f(R)$ gravity, it is possible to provide a unified
description of the late-time acceleration and of inflation as it
was demonstrated firstly in Ref. \cite{Nojiri:2003ft}. Also
generalized modified gravities, like for example Gauss-Bonnet
modified gravities
\cite{Nojiri:2005vv,Nojiri:2005jg,Cognola:2006eg,Koivisto:2006xf,Nojiri:2007te,Li:2007jm,Elizalde:2010jx,Bamba:2009uf},
teleparallel gravity
\cite{Cai:2015emx,Iorio:2012cm,Tamanini:2012hg,Ferraro:2011ks,Daouda:2011rt,Cai:2011tc,Dent:2011zz,Chen:2010va,Obukhov:2002tm}
and extensions of these
\cite{Harko:2011kv,Houndjo:2012ij,Alvarenga:2013syu}, can also
describe a plethora of cosmological and astrophysical phenomena.
Now the question is, which theory provides the most correct
description of our Universe, and this question can be answered
only by confronting each modified gravity with the observational
data. It is possible that the answer is simple, however most of
the theories can be compatible with the observations, so in
principle, all possible modified gravities should be scrutinized
in order to reveal the phenomenology these suggest.

In this line of research, recently the $f(R)$ gravity framework
was extended to include first and higher derivatives of the Ricci
scalar in the $f(R)$ gravity action \cite{Naruko:2015zze}. It was
shown that the resulting theory is free from ghost, under an
appropriate choice of the functional form of the Lagrangian. Later
on several studies in this framework were performed, see for
example
\cite{Yoshida:2017swb,Chervon:2017stf,Saridakis:2016ahq,Harko:2016xip,Otalora:2016dxe,Saitou:2016lvb,Saridakis:2016mjd,Akama:2017jsa,Tanahashi:2017kgn}.

In the present work, we shall investigate an extended $f(R)$
gravity model which contains first derivatives of the Ricci scalar
in the standard $f(R)$ gravity action. We shall derive the
gravitational field equations and we shall emphasize to
cosmological applications of the model at hand. Particularly, we
firstly demonstrate in a formal way how to obtain the
gravitational field equations, and we introduce appropriate
variables in order to cast the field equations in a convenient and
compact way. We specialize the field equations for a spatially
flat spacetime metric, and we study how it is possible to realize
various cosmological evolutions. As we demonstrate, the
realization of specific cosmological evolutions results to a
system of ordinary linear coupled differential equations, and the
general case can be quite tedious, so we focus our study on
specific limiting cases of the theory, which have some physical
significance. Also we demonstrate that the field equations can be
used as a reconstruction method, in which by providing the
cosmological evolution and the $f(R)$ gravity (or the kinetic term
related $X(R)$ function) it is possible to find which $X(R)$ (or
$f(R)$) gravity, may realize such an evolution. Interestingly
enough, it is possible to realize several cosmological evolutions
even in the case $f(R)=R$, such as bouncing cosmologies and even
inflationary evolutions of some sort, which in standard vacuum
$f(R)$ gravity was possible to realize for only specific forms of
the $f(R)$ gravity. Finally, we shall demonstrate that the kinetic
Ricci modified $f(R)$ gravity is equivalent with a multi-tensorial
scalar-tensor theory of gravity.

This paper is organized as follows: In section II we present the
theoretical framework of kinetic scalar curvature-corrected $f(R)$
gravity, and we derive the gravitational equations for a general
metric. In section III, we focus our study on cosmological
applications, by using a spatially flat metric. We investigate
various cosmological evolutions of physical interest and we find
the approximate form of the kinetic scalar curvature-corrected
$f(R)$ gravity which may realize such an evolution. We also show
how to treat the theory at hand as a reconstruction method for
realizing various cosmological evolutions, by specifying the
$f(R)$ gravity and the cosmological scale factor, and then finding
the function $X(R)$ of the kinetic term that realizes the given
cosmological scenario. Finally, in section IV we demonstrate that
the kinetic Ricci modified $f(R)$ gravity is equivalent with a
multi-tensorial scalar-tensor theory of gravity.

\section{$f(R)$ Gravity with Kinetic Scalar Curvature}

In this section we shall present the general formalism of kinetic
scalar curvature extended $f(R)$ gravity, and we shall derive the
field equations of this modified gravity. The action of the $f(R)$
gravity with kinetic scalar curvature term is the following,
\begin{equation}\label{act-1}
    \mathcal{S}=\int d^4x\sqrt{-g}\left(X(R) R_{,\sigma}R^{,\sigma}+f(R)\right)+\mathcal{S}_{matter}\, .
\end{equation}
where $X(R)$ and $f(R)$ are differentiable functions of the Ricci
scalar, and $\mathcal{S}_{matter}$ stands for the action of the
matter fluids present. The deviation from the standard $f(R)$
gravity action is obvious, and it is quantified by the term $\sim
X(R) R_{,\sigma}R^{,\sigma}$, which justifies the terminology
kinetic, since this is simply a kinetic term for the scalar
curvature $R$. A similar action to the above, in the context of
the Lagrange multipliers formalism
\cite{Capozziello:2013xn,Gao:2010gj,Cai:2010zma,Capozziello:2013hca,Makarenko:2016jsy},
was given in Ref. \cite{Capozziello:2013hca}. Upon variation of
the action (\ref{act-1}) with respect to the metric tensor $g^{\mu
\nu}$, we obtain the following,
\begin{multline}\label{var-act}
        \delta S = \int d^4x \sqrt{-g}\delta g^{\mu\nu}[-\frac{1}{2}g_{\mu\nu}( R_{,\sigma}R^{,\sigma}X(R) + f(R)) + X'(R)R_{\mu\nu}R_{,\sigma}R^{,\sigma} + \\+X(R)R_{,\nu}R_{,\mu} + f'(R)R_{\mu\nu} + (g_{\mu\nu}\Box - \nabla_{\mu}\nabla_{\nu})(R_{,\sigma}R^{,\sigma}X'+ f')\\ -2\nabla^{\sigma}X\nabla_{\sigma}RR_{\mu\nu} - 2X\Box R R_{\mu\nu} - 2(g_{\mu\nu}\Box - \nabla_{\mu}\nabla_{\nu})(\nabla^{\sigma}X\nabla_{\sigma}R+X\Box R)] \\+ \int d^4x\sqrt{-g}\nabla^{\alpha}M_\alpha -\int d^4x\sqrt{-g} \nabla_{\mu}L^\mu +2\int d^4x\sqrt{-g} \nabla^{\sigma}N_\sigma \\ +2\int d^4x\sqrt{-g} \nabla^{\sigma}P_\sigma - 2\int d^4x\sqrt{-g} \nabla^{\alpha}V_\alpha -2\int  d^4x\sqrt{-g} \nabla^{\sigma}W_\sigma + 2\int d^4x\sqrt{-g} \nabla_{\mu}Z^\mu \, ,
\end{multline}
where we introduced the tensorial quantities $M_\alpha$, $L^\mu$, $N^\sigma$, $P_\sigma$, $V_\alpha$, $W_\sigma$, $Z^\mu$, which are defined as follows,
\begin{equation}
 M_\alpha = \left(R_{,\sigma}R^{,\sigma}X'+ f'\right)g_{\beta\gamma}\nabla_{\alpha}\delta g^{\beta\gamma} - \nabla_{\alpha}\left(R_{,\sigma}R^{,\sigma}X'+ f'\right)g_{\beta\gamma}\delta g^{\beta\gamma}\, ,
\end{equation}
\begin{equation}
L^\mu = \left(R_{,\sigma}R^{,\sigma}X'+ f'\right) \nabla_{\beta}\delta g^{\mu\beta} - \nabla_{\beta}\left(R_{,\sigma}R^{,\sigma}X'+ f'\right) \delta g^{\mu\beta}\, ,
\end{equation}
\begin{equation}
N_\sigma = X\nabla_{\sigma}RR_{\mu\nu}\delta g^{\mu\nu}\, ,
\end{equation}
\begin{equation}
P_\sigma = XR_{,\sigma}g_{\alpha\beta}\Box\delta g^{\alpha\beta}\, ,
\end{equation}
\begin{equation}
V_\alpha = \left(\nabla^{\sigma}X\nabla_{\sigma}R + X\Box R\right)g_{\beta\gamma}\nabla_{\alpha}\delta g^{\beta\gamma} - \nabla_{\alpha}\left(\nabla^{\sigma}X\nabla_{\sigma}R + X\Box R\right)g_{\beta\gamma}\delta g^{\beta\gamma}\, ,
\end{equation}
\begin{equation}
W_\sigma = X\nabla_{\sigma}R\nabla_{\mu}\nabla_{\beta}\delta g^{\mu\beta}\, ,
\end{equation}
\begin{equation}
Z^\mu = \left(\nabla^{\sigma}X\nabla_{\sigma}R+X\Box R\right)\nabla_{\beta}\delta g^{\mu\beta} - \nabla_{\beta}\left(\nabla^{\sigma}X\nabla_{\sigma}R+X\Box R\right)\delta g^{\mu\beta} \, .
\end{equation}
By taking into account the fact that the 4-divergence terms lead
to three dimensional hypersurface integrals, that according to
Gauss-Stokes theorem tend to zero as surface terms, the vacuum
gravitational field equations are the following,
\begin{multline}\label{10}
-\frac{1}{2}g_{\mu\nu}f+R_{\mu\nu}f^{\prime}+D_{\mu\nu}f^{\prime}+(-\frac{1}{2}g_{\mu\nu}R_{, \sigma}R^{, \sigma}+R_{, \nu}R_{, \mu}-2R_{\mu\nu}\Box R)X+\\+(-R_{, \sigma}R^{, \sigma}R_{\mu\nu})X^{\prime}-2D_{\mu\nu}(\Box RX)-D_{\mu\nu}(R_{, \sigma}R^{, \sigma}X^{\prime})=0,
\end{multline}
where $D_{\mu\nu}=g_{\mu\nu}\Box - \nabla_{\mu} \nabla_{\nu}$. After some algebra, the field equations (\ref{10}) become,
\begin{equation}\label{endeq}
A^\mu_{\nu}X+B^\mu_{\nu}X'+C^\mu_{\nu}X''+F^\mu_{\nu}X'''-\dfrac{1}{2}\delta^\mu_{\nu}f+f'R_\nu^\mu+D^\mu_{\nu}f'=0\, ,
\end{equation}
where we introduced for notational convenience the tensorial quantities $A^\mu_\nu$, $B^\mu_\nu$, $C^\mu_\nu$, $F^{\mu}_\nu$, which are defined as follows,
\begin{equation}\label{cfs-1}
        A^\mu_\nu=R_{,\nu}R^{,\mu} - \frac{1}{2}\delta^\mu_\nu R_{,\sigma}R^{,\sigma}-2\Box RR^\mu_\nu-2D^\mu_\nu\Box R\, ,
    \end{equation}
    \begin{equation}\label{cfs-2}
        B^\mu_\nu=-R^\mu_{\nu}R_{,\sigma}R^{,\sigma}-D^\mu_\nu(R_{,\sigma}R^{,\sigma})-2\Box R D^\mu_\nu R \, ,
    \end{equation}
    \begin{equation}\label{cfs-3}
        C^\mu_\nu=-R_{,\sigma}R^{,\sigma}D^\mu_\nu R-2\Box R(\delta ^\mu_{\nu}R_{,\sigma}R^{,\sigma}-R_{,\mu}R^{,\nu}) \, ,
    \end{equation}
    \begin{equation}\label{cfs-4}
        F^{\mu}_\nu=-R_{,\sigma}R^{,\sigma}(\delta ^\mu_{\nu}R_{,\sigma}R^{,\sigma}-R^{,\mu} R_{,\nu}) \, .
    \end{equation}
Obviously, by setting $X(R)=0$, we recover the standard vacuum $f(R)$ gravity field equations,
\begin{equation}
        -\frac{1}{2}g_{\mu\nu}f(R) + f'(R)R_{\mu\nu} + (g_{\mu\nu}\Box - \nabla_{\mu}\nabla_{\nu})f' = 0\, ,
\end{equation}
where the prime in all the above equations denotes differentiation
with respect to the Ricci scalar. Thus we derived the
gravitational equation of the $f(R)$ gravity with kinetic scalar
curvature terms, for the vacuum case of the action (\ref{act-1}),
which are Eqs. (\ref{endeq}) along with the definitions of the
tensorial quantities $A^\mu_\nu$, $B^\mu_\nu$, $C^\mu_\nu$,
$F^{\mu}_\nu$, namely Eqs. (\ref{cfs-1})-(\ref{cfs-4}).

It is useful to present the trace of the gravitational field
equations (\ref{endeq}). Performing a contraction of the indices,
we obtain the following equation,
\begin{equation}\label{trace}
    AX + BX' + CX'' + FX''' -2f +Rf' + 3\Box f' = 0
\end{equation}
where the parameters $A$, $B$, $C$, $F$, are defined as follows,
\begin{equation}
    A = -\hat{R} - 2R\Box R - 6\Box^2R,~~\hat{R} = R_{,\sigma}R^{,\sigma}\, ,
    \label{Aii}
\end{equation}
\begin{equation}
    B = -R\hat{R} - 3\Box(\hat{R}) -6\left( \Box R \right)^2\, ,
    \label{Bii}
\end{equation}
\begin{equation}
    C = -9\hat{R}\Box{R}\, ,
    \label{Cii}
\end{equation}
\begin{equation}
    F = -3\left( \hat{R} \right)^2 \, .
    \label{Fii}
\end{equation}
For cosmological applications in this paper we shall consider a spatially flat Friedmann-Robertson-Walker (FRW) Universe with metric,
\begin{equation} \label{frw}
  ds^2=-dt^2+a^2(t)\left( dr^2 + r^2\left( d\theta^2 + \sin^2\theta  \right)d\phi^2 \right)\, ,
\end{equation}
where $a$ is the scale factor. For later purposes, it is worth
finding the functional form of the quantities $A$, $B$, $C$, $F$
appearing in Eqs. (\ref{Aii})-(\ref{Fii}) for a FRW Universe,
which are,
\begin{eqnarray}
A = 1728H^4 \dot H-288H^3\ddot H+(-576(\dot H)^2-1044\dddot H)H^2+(-3420\dot H \ddot H-360\ddddot H)H-
{} \\ \nonumber {}
-576(\dot H)^3-720\dot H\ddot H-504(\ddot H)^2-36\ddddot H
\label{Aii1}
\end{eqnarray}
\begin{eqnarray}
B = -3456(\dot H)^4-6912H^2(\dot H)^3+(6912H^4-18144H\ddot H-1728\dddot H)(\dot H)^2+
{} \\ \nonumber {}
(-6912H^3\ddot H-6048H^2\ddot H-864H \ddddot H-2376(\ddot H)^2)\dot H -
{} \\ \nonumber {}
-5616H^2(\ddot H)^2-3240H\ddot H\dddot H-216\ddot H\ddddot H-216(\dddot H)^2
\label{Bii1}
\end{eqnarray}
\begin{equation}
C = -1944(4H\dot H+\ddot H)^2(12H^2\dot H+7H\ddot H+4(\dot H)^2+\ddot H)
\label{Cii1}
\end{equation}
\begin{equation}
F = -3888(4H\dot H+\ddot H)^4
\label{Fii1}
\end{equation}
where $H$ stands for the Hubble rate of the Universe
$H=\frac{\dot{a}}{a}$. The above Eqs. (\ref{Aii1})-(\ref{Fii1})
will be useful later on, when we discuss cosmological solutions.

Also it will be convenient to divide the general equation (\ref{endeq}) in two parts,
\begin{equation}\label{eq22}
    Q^\mu_\nu + L^\mu_\nu = 0
\end{equation}
where $Q^\mu_\nu$  is the part containing the function $X(R)$ and its derivatives,
\begin{equation}
Q^\mu_\nu=A^\mu_{\nu}X+B^\mu_{\nu}X'+C^\mu_{\nu}X''+F^\mu_{\nu}X'''\, ,  \label{endeq2}
\end{equation}
the trace of which is equal to,
\begin{equation}
    Q = AX + BX' + CX'' + FX'''\, ,
    \label{Gii}
\end{equation}
and $L^\mu_k$ is the part containing the function $f(R)$ and its derivatives,
\begin{equation}
 L^\mu_\nu= -\dfrac{1}{2}\delta^\mu_\nu f+f^{'}R^\mu_\nu+D^\mu_\nu(f^{'})\, , \label{Lik}
\end{equation}
with the corresponding trace being in this case,
\begin{equation}
    L={L}_\mu^\mu = -2f + Rf' + 3\Box f'\, .
    \label{Gbarii}
\end{equation}
Having described the general formalism of the kinetic Ricci
modified $f(R)$ gravity, in the next sections we shall consider
several cosmological realizations of this theory, focusing on
inflationary evolutions mainly. Also we demonstrate the
equivalence of the theory with a scalar-tensor theory at some
later section. Before we continue, we need to briefly discuss a
somewhat critical issue regarding the kinetic Ricci modified
$f(R)$ gravity, having to do with the stability of the theory. All
the cosmological realizations of this formalism, which we present
in the next section, should be examined towards their stability.
According to Ref. \cite{Naruko:2015zze}, ghost instabilities are
absent, since the kinetic terms are appropriately chosen, however
a potential source of instability in this theory can be traced in
Eq. (2.17) of Ref. \cite{Naruko:2015zze}, namely in the potential,
which is unbounded from below since it is linear in the scalar
field. This issue in turn could make the theory less
self-consistent, due to the severe instabilities caused. This
issue should be appropriately examined in a future study, which is
beyond the scopes of this demonstrational version of our work.

\section{Cosmology with $f(R,(\nabla R)^2)$ Gravity}

Let us now focus on cosmological applications of the kinetic Ricci
modified $f(R)$ gravity, and we discuss the general framework of
the theory in the context of a flat FRW background. The non-zero
components of the Ricci tensor for the FRW metric are,
\begin{align*}
    R_0^0 &=  3\frac{\ddot{a}}{a} \\
    R_1^1 = R_2^2 = R_3^3 &= \frac{\ddot{a}}{a} + 2\frac{\dot{a}^2}{a^2}\, ,
\end{align*}
and the corresponding Ricci scalar is,
\begin{equation}\label{riccisk}
  R =  6\left(\frac{\dot{a}^2}{a^2}+\frac{\ddot{a}}{a}\right)\, ,
\end{equation}
Using the equation (\ref{endeq}),
\begin{equation}\nonumber
    A^\mu_\nu X + B^\mu_\nu X' + C^\mu_\nu X'' + F^{\mu}_\nu X''' - \frac{1}{2}\delta^\mu_\nu f + f'R^\mu_\nu  + D^\mu_\nu (f') = 0
\end{equation}
we may calculate each component of it, for the metric (\ref{frw})
in terms of Hubble rate $H$, and the resulting set of the
components of $A^\mu_n$, $B^\mu_n$, $C^\mu_n$ and $F^\mu_n$ are
the following,
\begin{multline}\label{fAeq}
    A^0_0 = 18\left( 24H^4\dot{H} - 10H^3\ddot{H} - 32\dot{H}^2H^2 - 12\dddot{H}H^2\right. - \left. 24\ddot{H}H\dot{H} + 8\dot{H}^3 - 2H\ddddot{H} + 2\dot{H}\dddot{H} - \ddot{H}^2\right)\, ,
\end{multline}
\begin{multline}
    A^1_1=A^2_2=A^3_3 = 6\left( 72H^4 \dot{H} - 6H^3\ddot{H} - 46\dddot{H}H^2 - 166\ddot{H}H\dot{H} \right. - \left. 40\dot{H}^3 -18H\ddddot{H} - 42\dot{H}\dddot{H} - 27 \ddot{H}^2 - 2H^{(5)} \right)\, ,
\end{multline}
\begin{align}
    B^0_0 &= -1728\left( H\dddot{H} + \frac{1}{4}\left( 17H^2 - \dot{H} \right)\ddot{H} + H\dot{H}\left( H^2 + 3\dot{H} \right) \right)\left( H\dot{H} + \frac{1}{4}\ddot{H} \right)\\
    B^1_1=B^2_2=B^3_3 &= -576\left( H\dot{H} + \frac{1}{4}\ddot{H} \right)\left( \frac{1}{2}\ddddot{H} + \frac{11}{2}\dddot{H} + \left( \frac{57}{4}H^2 + \frac{23}{4}\dot{H} \right)\ddot{H} + H\dot{H}\left( H^2 + 13\dot{H} \right)\right)\, ,
    \label{fBeq}
\end{align}
\begin{align}
    C^0_0 &= -648H\left( 4\dot{H}H + \ddot{H} \right)^3\\
    C^1_1=C^2_2=C^3_3 &= -216\left( 4H\dot{H} + \ddot{H} \right)^2\left( 32\dot{H}H^2 + 20\ddot{H}H + 12\dot{H}^2 + 3\dddot{H} \right)\, ,
    \label{fCeq}
\end{align}
\begin{align}
    F^0_0 &= 0 \\
    F^1_1=F^2_2=F^3_3 &= -1296\left( 4H\dot{H}+\ddot{H} \right)^4\, .
    \label{fFeq}
\end{align}
The above components, will be useful for the study of specific
cosmological solutions, as we will shortly show. The kinetic Ricci
modified $f(R)$ gravity is a geometric theory, as the $f(R)$
gravity theory is too, so it is useful to rewrite the
gravitational field equations in a way so that the geometric
effects can be formed as a perfect fluid and the resulting
Einstein equations can be written in the standard way as in the
Einstein-Hilbert case. To this end, consider the $f(R)$ gravity
part tensor $L^\mu_\nu$ of Eq. (\ref{Lik}),
\begin{equation}
    L^\mu_\nu  = - \frac{1}{2}\delta^\mu_\nu f + f'R^\mu_\nu  + D^\mu_\nu (f')\, ,
\end{equation}
the non-zero components of which are,
\begin{eqnarray}
    L^0_0 = -3H^2f' + \frac{R f'-f}{2} - 3H\dot{R}f''\\
    L^1_1 =L^2_2=L^3_3= -\frac{1}{2}f + \left( \frac{R}{2} + 2\dot{H} + 3H^2 \right)f' + \left( \ddot{R} + 2H\dot{R} \right)f'' + \left( \dot{R} \right)^2f'''\, .
    \label{Lcoef}
\end{eqnarray}
In effect, we can rewrite Eq. (\ref{eq22}) in a perfect fluid form, by separating the terms $H^2$ and $2\dot{H}+3H^2$, as follows,
\begin{equation}
    H^2 = \frac{1}{3f'}\left( \kappa\rho + \frac{Rf'-f}{2} - 3H\dot{R}f'' +A^0_0X + B^0_0X' + C^0_0X'' \right)\, ,
    \label{friedman1}
\end{equation}
\begin{multline}
    2\dot{H} + 3H^2 = \frac{1}{f'}\left( -\kappa p + \frac{f-Rf'}{2} -\left( \ddot{R} + 2H\dot{R} \right)f'' - \left( \dot{R} \right)^2f''' \right. - \left. A^1_1X - B^1_1X' - C^1_1X'' - F^1_1X'''  \right)\, .
    \label{friedman1b}
\end{multline}
We can bring the above equations to the standard Friedmann
equations form of Einstein-Hilbert gravity, if we assume that a
perfect fluid originating from geometry has a stress tensor with
energy and pressure components defined as follows,
\begin{eqnarray}
    \rho_{eff} &=& \frac{1}{f'}\left(\frac{Rf'-f}{2} - 3H\dot{R}f'' +A^0_0X + B^0_0X' + C^0_0X''\right)\\
    p_{eff} &=& \frac{1}{f'}\left(\frac{Rf'-f}{2} + \left( \ddot{R} + 2H\dot{R} \right)f'' + \left( \dot{R} \right)^2f''' + A^1_1X + B^1_1X' + C^1_1X'' + F^1_1X'''  \right)\, .
    \label{effectiveprho}
\end{eqnarray}
Hence, in the absence of standard matter fluids, in which case $\rho=p=0$ in Eqs. (\ref{friedman1}) and (\ref{friedman1b}), we can rewrite these as follows,
\begin{eqnarray}
    H^2 &=& \frac{\rho_{eff}}{3}\, , \\
    2\dot{H} + 3H^2 &=& -p_{eff}\, .
    \label{friedmann2}
\end{eqnarray}
The same equations can be easily rewritten in form of Friedmann equations of standard Einstein-Hilbert gravity,
\begin{eqnarray}
    H^2 &=& \frac{\rho_{eff}}{3} \\
    \frac{\ddot{a}}{a} &=& -\frac{1}{6}\left( \rho_{eff} + 3p_{eff} \right)\, ,
    \label{friedman3}
\end{eqnarray}
and the contribution in $\rho_{eff}$ and $p_{eff}$ is purely
geometrically originating. Having the Friedmann equations at hand,
we can proceeding in considering several cosmological solutions in
the context of kinetic Ricci $f(R)$ gravity.

\subsection{Specific Cosmological Solutions}

Recall that the part of the field equations which contain the $X(R)$ contribution in Eq. (\ref{endeq2}), is,
\begin{equation}\nonumber
    Q^\mu_\nu = A^\mu_\nu X + B^\mu_\nu X' + C^\mu_\nu X'' + F^{\mu}_\nu X'''
\end{equation}
while the part containing the $f(R)$ contribution is the one appearing in Eq. (\ref{Lik}), which we quote here for reading convenience,
\begin{equation} \nonumber
    L^\mu_\nu = -\frac{1}{2}\delta^\mu_\nu f + f' R^\mu _\nu  + D^\mu_\nu(f')\, .
    \label{FRpart}
\end{equation}
Let us now consider some simple cosmological solutions, and we
start off our analysis with a toy model of a static Universe, in
which case $a(t)=C$, so the field equations (\ref{endeq2}) and
(\ref{Lik}), reduce both to,
\begin{eqnarray} \nonumber
    -\dfrac{f}{2}=0
\end{eqnarray}
Thus in this case we have $f(R)=0$, and $X(R)$ is an arbitrary
function. Consider now the case of a de Sitter cosmological
evolution, in which case $a(t) = e^{kt}$, where $k>0$. In this
case, the field equations take the following form,
\begin{eqnarray} \nonumber
-\dfrac{f}{2}+\dfrac{3}{4}k^2f^{'}=0\, ,
\end{eqnarray}
The solution of the above equation is $f=\mathcal{C}\exp(\frac{2
R}{3 k^2})$, where $\mathcal{C}$ is an arbitrary integration
constant. For the purely de Sitter evolution, all the coefficients
$A^i_i$, $B^{i}_i$, $C^i,_i$, $F^i_i$, $i=1,2$ are equal to zero,
so the field equations are satisfied for an arbitrary function
$X(R)$. Phenomenologically, this might be interesting for this
particular cosmological solution, due to the fact that the effects
of the $X(R)$ gravity are trivial, and therefore have no direct
effect on the slow-roll indices of inflation. However, the
situation might change if the evolution is a quasi de Sitter
evolution, as we show later on.

Let us now consider a power law evolution, with scale factor $a(t) = t^m$. We shall make use of the following formulas,
\begin{eqnarray}
    X' &=& \frac{\dot{X}}{\dot{R}}\, ,\\
    X'' &=& \frac{\ddot{X}(t)}{R'(t)^2}-\frac{\dot{X}(t) R''(t)}{R'(t)^3}\, ,\\
    X''' &=& \frac{X^{(3)}(t)}{R'(t)^2}-\frac{2 \ddot{X}(t) R''(t)}{R'(t)^3}-\frac{R^{(3)}(t) \dot{X}(t)}{R'(t)^3}+\frac{2 \dot{X}(t) R''(t)^2}{R'(t)^4}\, ,
    \label{helpXp}
\end{eqnarray}
which relate the derivatives of the function $X(R)$ with respect
to the Ricci scalar, denoted with a ``prime'', with the
derivatives of the function $X(t)$ with respect to the cosmic
time, denoted with a ``dot''. From (\ref{endeq}) we have,
\begin{multline}
    Q^0_0 = \frac{5832m^2\left( m - \frac{1}{2} \right)}{t^6}\left( -m^2\left( m - \frac{1}{2} \right)^2t^2\ddot{X}\right. +\\+ \left. m\left( \frac{11}{9} + m^2t + \left( -\frac{t}{2} - \frac{1}{9} \right)m\right)\left( m - \frac{1}{2} \right)t\dot{X} + \frac{2}{27}\left( m^2 +\frac{8}{3}m - \frac{10}{3} \right)X \right) \label{Q00X}\, ,
\end{multline}
\begin{multline}
    Q^1_1= Q^2_2=Q^3_3= \frac{17496m\left( m - \frac{1}{2} \right)}{t^6}\left( -m^3\left( m - \frac{1}{2} \right)^3t^3\dddot{X} \right. +\\ +  m^2\left( m - \frac{1}{2} \right)^2\left( 1 + m^2t + \left( -\frac{t}{2} - \frac{8}{9} \right)m \right)t^2\ddot{X} +\\+ \frac{8}{9}m\left( m - \frac{1}{2} \right)\left( -\frac{1}{3} + m^3t + \left( -\frac{13t}{8} - \frac{1}{72} \right)m^2  + \left( \frac{9t}{16} + \frac{41}{72} \right)m \right)t\dot{X} +\\+\left. \frac{2}{81}\left( m - 2 \right)\left( m^2 + \frac{8}{3}m - \frac{10}{3} \right)X \right)
    \label{eq_dustsol}
\end{multline}
From a first observation, it seems that for certain values of the
parameter $m$, significant simplifications occur in the field
equations, which we consider now. For example, the case $m=0$ is
the static Universe toy model we discussed earlier. More
significantly, the value $m=\frac{1}{2}$ simplifies considerably
the field equations, and in this case we have,
\begin{eqnarray} \nonumber\label{asx}
-\dfrac{f}{2}-\dfrac{3}{4}f^{'}=0
\\-\dfrac{f}{2}+\dfrac{1}{4}f^{'}=0\nonumber
\end{eqnarray}
In this case too, all the coefficients $A^i_i$, $B^{i}_i$,
$C^i,_i$, $F^i_i$, $i=1,2$ are equal to zero as it can be checked,
so the function $X(R)$ is arbitrary, and also the solution of the
differential equations (\ref{asx}) is, $f(R)=\mathcal{C}e^{2R}$,
where $\mathcal{C}$ is an integration constant.

Let us now assume that the function $X(R)$ has a simple form,
$X(R)=$const. In this case, all derivatives of $X(R)$ with respect
to the Ricci scalar are equal to zero. Also for $m=-\dfrac{4}{3}+
\dfrac{\sqrt{46}}{3}$, the $X$-dependent part of the field
equations vanish. For general $f(R)$ dependent part,
\begin{multline}
L^0_0=-\dfrac{f}{2}+\left(\dfrac{\left(-22\sqrt{46}+148\right)t}{8\left(-136+19\sqrt{46}\right)}-\dfrac{3\left(1696\sqrt{46}-11344\right)t^2}{32\left(-136+19\sqrt{46}\right)^2}\right)\dot f+\dfrac{3\left(1696\sqrt{46}-11344\right)t^2}{32\left(-136+19\sqrt{46}\right)^2}\ddot f=0\, ,
\end{multline}
\begin{multline}
L^1_1=L^2_2=L^3_3=-\dfrac{f}{2}+\left(\dfrac{\left(-54\sqrt{46}+396\right)t}{24\left(-136+19\sqrt{46}\right)}-\dfrac{\left(4760\sqrt{46}-32480\right)t^2}{32\left(-136+19\sqrt{46}\right)^2}\right)\dot f+\\+\left(\dfrac{\left(4760\sqrt{46}-32480\right)t^2}{32\left(-136+19\sqrt{46}\right)^2}+\dfrac{\left(-69768\sqrt{46}++473877\right)t^3}{18\left(-136+19\sqrt{46}\right)^3}\right)\ddot f+\dfrac{\left(69768\sqrt{46}-473877\right)t^3}{18\left(-136+19\sqrt{46}\right)^3}\dddot f=0\, ,
\end{multline}
which can be solved with respect to $f(t)$ if the Hubble rate of
the cosmological evolution is given. This technique is the core of
a general reconstruction method for kinetic Ricci $f(R)$ gravity,
which we shall present in the next section.

As a final example, let us consider a quasi-de Sitter evolution,
with scale factor $a(t)=e^{H_\ast t-\epsilon t^2}$, where
$\epsilon$ is assumed to take small values. Note that this case of
quasi-de Sitter evolution is different from a phenomenological
point of view from an inflationary quasi-de Sitter evolution
$a(t)=e^{H_0t-H_i t^2}$, in which case the parameters $H_0$ and
$H_i$ are constrained by observations for the ordinary Starobinsky
model \cite{starobinsky}, see for example Ref.
\cite{Odintsov:2015gba} for a concrete analysis of the parameter
space. Coming to the problem at hand, by assuming that terms $\sim
\epsilon^2$ and higher orders tend to zero, obtain
\begin{equation}
    H = H_* - 2\epsilon t;\ \ \ \dot{H} = -2\epsilon; \ \ \ \dot{H}^2 \approx 0; \ \ \ \frac{\ddot{a}}{a} = -2\epsilon + \left( H_* - 2\epsilon t \right)^2; \ \ \ R \approx 12 \left( H_*^2 - \left( 4H_*t + 1  \right)\epsilon \right)\, .
    \label{eq_anc3}
\end{equation}
Upon substitution \eqref{eq_anc3} into \eqref{endeq} the coefficients \eqref{endeq} become,
\begin{equation}
    A_\mu^\nu \approx -864\epsilon H_*^4;\ \ \ B_\mu^\nu\approx 0;\ \ \ C_\mu^\nu\approx 0;\ \ \ F_\mu^\nu \approx 0\, .
    \label{eq_anc5}
\end{equation}
As a result we obtain the following field equations,
\begin{equation}
    288\epsilon H_*^2 f'' + \left( \frac{R}{2} - 6\epsilon \right)f' - 1728\epsilon H_*^4X-f= 0\, ,
    \label{eq_mt1}
\end{equation}
\begin{equation}
    192\epsilon H_*^2 f'' + \left( \frac{R}{2} + 2\epsilon \right)f' - 1728\epsilon H_*^4 X - f = 0\, .
    \label{eq_mt2}
\end{equation}
From (\ref{eq_mt1}) and (\ref{eq_mt2}), we can find the approximate forms of $f(R)$ and of $X(R)$, which are,
\begin{equation}
    f(R) = 1728\epsilon H_*^4\left( 24 C_2 H_*^2 \exp\left(\frac{R}{12H_*^2}\right) - C_1 \right)\, ,
    \label{eq_mt3}
\end{equation}
\begin{equation}
    X(R) = C_1 + C_2 \exp\left(\frac{R}{12H_*^2}\right)\left( R - 24H_*^2 + 36\epsilon \right)\, .
    \label{eq_mt4}
\end{equation}

The solutions \eqref{eq_mt3} and \eqref{eq_mt4} correspond to the
model with quasi-de Sitter evolution. In the next section, we also
discuss how the model at hand actually constitutes a
reconstruction method, in the context of which, several
cosmological evolutions can be realized.

\subsection{A General Reconstruction Technique for Kinetic Ricci Extended $f(R)$ Gravity}

Essentially, the field equations (\ref{friedmann2}),
\eqref{friedman3} with the definitions \eqref{effectiveprho}, can
be used as a reconstruction method in which, given the Hubble rate
and one of the functions $f(R)$ and $X(R)$, it is possible to
determine the other function, and hence determine the kinetic
Ricci modified $f(R)$ gravity which realizes the given
cosmological evolution. Apparently, the most interesting case is
by choosing $f(R)=R$, so the theory is standard Einstein-Hilbert
gravity with scalar curvature kinetic corrections. In the
following we shall investigate this case, however other choices
for the function $f(R)$ are possible.

Let us demonstrate explicitly how the reconstruction method works,
by using some well known cosmological evolutions, starting with
the symmetric bounce case, in which case the scale factor and the
Hubble rate are,
\begin{equation}\label{symmetricbounce}
a(t)=e^{\beta t^2},\,\,\,H(t)=2 \beta t\, ,
\end{equation}
and since $a(0)\neq 0$ this is a non-singular bounce. Clearly the
bouncing behavior is acquired from the change in the sign of the
Hubble rate before and after the bouncing point $t=0$. In the
literature, there are standard works that study this type of
cosmological evolution, in the context of $f(R)$ gravity
\cite{Odintsov:2015gba}. To start with, assume that $f(R)=R$, so
the standard general relativity is assumed in the $f(R)$ part, and
the corresponding vacuum action is,
\begin{equation}\label{symmetricbounceaction}
    \mathcal{S}=\int d^4x\sqrt{-g}\left(R+X(R) R_{,\sigma}R^{,\sigma}\right)\, .
\end{equation}
In this case, the non-zero components of the tensorial quantities $A^\mu_n$, $B^\mu_n$, $C^\mu_n$ and $F^\mu_n$ given in Eqs. \eqref{fAeq}-\eqref{fFeq} are,
\begin{equation}\label{fAeqnew}
    A^0_0 = 18 \left(64 \beta ^3+768 \beta ^5 t^4-512 \beta ^4 t^2\right)\, ,
\end{equation}
\begin{equation}
    A^1_1=A^2_2=A^3_3 = 6 \left(2304 \beta ^5 t^4-320 \beta ^3\right)\, ,
\end{equation}
\begin{align}
    B^0_0 &= -27648 \beta ^4 t^2 \left(6 \beta +4 \beta ^2 t^2\right)\, ,\\
    B^1_1=B^2_2=B^3_3 &= -36864 \beta ^6 t^4-239616 \beta ^5 t^2\, ,
    \label{fBeqnew}
\end{align}
\begin{align}
    C^0_0 &= -5308416 \beta ^7 t^4\, ,\\
    C^1_1=C^2_2=C^3_3 &= -331776 \beta ^5 t^3 \left(48 \beta ^2+256 \beta ^3 t^2\right))\, ,
    \label{fCeqnew}
\end{align}
\begin{align}
    F^0_0 &= 0 \\
    F^1_1=F^2_2=F^3_3 &= 331776 \beta ^4 t^2\, .
    \label{fFeqnew}
\end{align}
It is conceivable that even in the case $f(R)=R$, the resulting
differential equation for the function $X(t)$ is impossible to
solve analytically. So we shall focus our analysis for cosmic
times near the bouncing point, so for $t\sim 0$. By making this
assumption, we shall keep only the lowest powers of the cosmic
time dependent terms, and the resulting approximate differential
equation for $X(t)$ (\ref{friedman1}) is,
\begin{equation}\label{diffeqnsforxt}
-576 \beta ^3 t^2\ddot{X}(t)-1152 \beta ^3 t\dot{X}(t)+1152 \beta ^3 X(t)-4 \beta ^2 t^2=0\, ,
\end{equation}
which can be solved analytically, and the resulting solution is,
\begin{equation}\label{solutionforxt}
X(t)=-\frac{t^2}{576 \beta }+\frac{\Lambda_a}{t^2}+\Lambda_b t\, ,
\end{equation}
where $\Lambda_a$ and $\Lambda_b$ are integration constants. Also
the Ricci scalar for the symmetric bounce (\ref{symmetricbounce})
is equal to $R(t)=12 \beta +48 \beta ^2 t^2$, so by inverting the
function $R(t)$ we obtain two solutions $t(R)=\pm \frac{\sqrt{R-12
\beta }}{4 \sqrt{3} \beta }$. Finally, substituting $t(R)$ in the
solution (\ref{solutionforxt}), we obtain the following two
approximate solutions near the bouncing point,
\begin{align}\label{finalrsolutionsforxr}
& X(R)\simeq -\frac{R-12 \beta }{27648 \beta ^3}+\frac{48 \beta ^2 \Lambda_b}{R-12 \beta }+\frac{\Lambda_a \sqrt{R-12 \beta }}{4 \sqrt{3} \beta }\, ,\\ \notag &
X(R)\simeq -\frac{R-12 \beta }{27648 \beta ^3}+\frac{48 \beta ^2 \Lambda_b}{R-12 \beta }-\frac{\Lambda_a \sqrt{R-12 \beta }}{4 \sqrt{3} \beta }\, .
\end{align}
In both cases the solutions are constrained for values of the
curvature $R>12 \beta$. Let us now consider another case, so we
choose $f(R)=R+\frac{1}{36 Hi}R^2$, so the $f(R)$ part is the
Starobinsky model \cite{starobinsky}, and the corresponding action
is,
\begin{equation}\label{symmetricbounceaction1}
    \mathcal{S}=\int d^4x\sqrt{-g}\left(R+\frac{1}{36 Hi}R^2+X(R) R_{,\sigma}R^{,\sigma}\right)\, .
\end{equation}
Let us now investigate which $X(R)$ function realizes the
symmetric bounce cosmology (\ref{symmetricbounce}), emphasizing
near the bouncing point. By following the steps of the previous
case, the resulting differential equation at leading order near
the bouncing point is,
\begin{equation}\label{diffeqnsforxtnew}
-55296 \beta ^5 t^3\ddot{X}(t)-110592 \beta ^5 t^2\dot{X}(t)+110592 \beta ^5 tX(t)+192 \beta ^4 t=0\, ,
\end{equation}
which can be solved analytically, and the solution is,
\begin{equation}\label{solutionstarobinsky}
X(t)=-\frac{1}{576 \beta }+\frac{\Lambda_b}{t^2}+\Lambda_a t\, ,
\end{equation}
where $\Lambda_a$ and $\Lambda_b$ are integration constants. By substituting the function $t(R)=\pm \frac{\sqrt{R-12 \beta }}{4 \sqrt{3} \beta }$ in Eq. (\ref{solutionstarobinsky}), we find the following two solutions for $X(R)$,
\begin{align}\label{finalrsolutionsforxrnew}
& X(R)\simeq -\frac{1}{576 \beta }+\frac{48 \beta ^2 \Lambda_b}{R-12 \beta }+\frac{\Lambda_a \sqrt{R-12 \beta }}{4 \sqrt{3} \beta }\, ,\\ \notag &
X(R)\simeq -\frac{1}{576 \beta }+\frac{48 \beta ^2 \Lambda_b}{R-12 \beta }-\frac{\Lambda_a \sqrt{R-12 \beta }}{4 \sqrt{3} \beta }\, ,
\end{align}
so in this case too, the condition $R>12\beta$ is required for the consistency of the solution.

As a final example, let us consider the quasi-de Sitter evolution with scale factor and Hubble rate,
\begin{equation}\label{quasidesitternew}
a(t)=e^{H_0t-H_it^2},\,\,\,H(t)=H_0-2H_i t\, .
\end{equation}
The quasi-de Sitter evolution (\ref{quasidesitternew}) is known to
be realized by the Starobinsky $R^2$ model, and the parameters
$H_0$ and $H_i$ are constrained from the Planck data, see Ref.
\cite{Odintsov:2015gba} for a thorough analysis of the parameter
space. Instead of using the $R^2$ model, we shall assume that
$f(R)=R$, and we shall find which kinetic Ricci $X(R)$ gravity can
realize this inflationary evolution, emphasizing to small cosmic
times. The action is again given by Eq.
(\ref{symmetricbounceaction}), and in this case, the non-zero
components of the tensorial quantities $A^\mu_n$, $B^\mu_n$,
$C^\mu_n$ and $F^\mu_n$ given in Eqs. \eqref{fAeq}-\eqref{fFeq},
at leading order in the cosmic time $t$, are,
\begin{align}\label{fAeqnewnewnew}
    & A^0_0 = -864 H_0^4 H_i+6912 H_0^3 H_i^2 t-2304 H_0^2 H_i^2+9216 H_0 H_i^3 t-1152 H_i^3\, ,\\ \notag &
    A^1_1=A^2_2=A^3_3 = -864 H_0^4 H_i+6912 H_0^3 H_i^2 t+1920 H_i^3\, ,\\ \notag &
    B^0_0 = -6912 H_0^4 H_i^2+55296 H_0^3 H_i^3 t+41472 H_0^2 H_i^3-165888 H_0 H_i^4 t\, ,\\ \notag &
    B^1_1=B^2_2=B^3_3= -2304 H_0^4 H_i^2+18432 H_0^3 H_i^3 t+59904 H_0^2 H_i^3-239616 H_0 H_i^4 t\, , \\ \notag &
    C^0_0= 331776 H_0^4 H_i^3-2654208 H_0^3 H_i^4 t\, ,\\ \notag &
    C^1_1=C^2_2=C^3_3 = 2654208 H_0^5 H_i^3-26542080 H_0^4 H_i^4 t-1990656 H_0^3 H_i^4+11943936 H_0^2 H_i^5 t\, , \\ \notag &
    F^0_0 = 0\, , \\ \notag &
    F^1_1=F^2_2=F^3_3 = 82944 H_0^2 H_i^2-331776 H_0 H_i^3 t\, .
\end{align}
Then, it is possible to obtain the differential equation which
will yield the function $X(t)$ at leading order, by using  Eq.
(\ref{friedman1}), which becomes in this case,
\begin{equation}\label{diffeqnsforxtnewnew}
144 H_0^2 H_i\ddot{X}(t)+\left(144 H_0^3 H_i-864 H_0 H_i^2\right)\dot{X}(t)-\left( 864 H_0^4 H_i+2304 H_0^2 H_i^2+1152 H_i^3\right) X(t)-H_0^2=0\, ,
\end{equation}
which can be solved analytically, and the solution is in this case,
\begin{equation}\label{solutionforxtnewnew}
X(t)=\mathcal{C}_a e^{\mu  t}+\mathcal{C}_b e^{-\nu  t}+\Lambda\, ,
\end{equation}
where $\Lambda$, $\mathcal{C}_a$ and $\mathcal{C}_b$ are integration constants, and the parameters $\mu$ and $\nu$ are equal to,
\begin{equation}\label{munudefinitions}
\mu=\frac{-H_0^3+\sqrt{25 H_0^6+52 H_0^4 H_i+68 H_0^2 H_i^2}+6 H_0 H_i}{2 H_0^2},\,\,\, \nu=\frac{H_0^3+\sqrt{25 H_0^6+52 H_0^4 H_i+68 H_0^2 H_i^2}-6 H_0 H_i}{2 H_0^2}\, .
\end{equation}
The Ricci scalar for the quasi-de Sitter evolution at hand is at
leading order $R(t)=12 H_0^2-48 H_0 H_i t-12 H_i$, so by inverting
this function, we obtain $t(R)=\frac{12 H_0^2-12 H_i-r}{48 H_0
H_i}$. So substituting $t(R)$ in $X(t)$
(\ref{solutionforxtnewnew}), we obtain the solution,
\begin{align}\label{finalrsolutionsforxrnewnew}
& X(R)\simeq \mathcal{C}_a e^{\frac{\mu  \left(12 H_0^2-12 H_i-R\right)}{48 H_0 H_i}}+\mathcal{C}_b e^{-\frac{\nu  \left(12 H_0^2-12 H_i-R\right)}{48 H_0 H_i}}+\Lambda\, .
\end{align}
Hence, the inflationary quasi-de Sitter evolution
(\ref{quasidesitternew}), which is realized by the vacuum $R^2$
model, can be realized by standard Einstein-Hilbert gravity with
exponential kinetic corrections of the scalar curvature, of the
form (\ref{finalrsolutionsforxrnewnew}).

\section{Scalar-multi-tensorial Equivalence}

The kinetic Ricci $f(R)$ gravity we considered in the previous
sections, belongs to a general class of models of the form $ f(R,
\nabla_\mu R, ,..., \nabla^n R)$, studied in Ref.
\cite{Cuzinatto:2016ehv}.  In this section we shall demonstrate
the equivalence of the theory at hand with a scalar-tensor theory.
To this end we can define the functional dependence
\begin{equation}
    \Phi(R,R_{,\mu}) = f(R) + X(R)R_{,\mu}R^{,\nu}\, ,
    \label{Phidef}
\end{equation}
and the action \eqref{act-1} is written in the following form,
\begin{equation}
    S = \int d^4x \sqrt{-g}\Phi(R,R_{,\mu}) +S_{matter}\, .
    \label{act2}
\end{equation}
Following the formalism firstly proposed in Ref. \cite{Cuzinatto:2016ehv}, we make the following substitutions,
we can use the substitution
\begin{eqnarray}
    \xi &=& R\\
    \xi_{\mu} &=& R_{,\mu}\, ,
    \label{subst}
\end{eqnarray}
Applying the above, the function \eqref{Phidef} transforms as follows,
\begin{equation}
    \Phi(\xi,\xi_{\mu}) = f(\xi) + X(\xi) \xi_{\mu}\xi^{\mu}\, .
    \label{Phidef2}
\end{equation}
The first order derivatives are equal to,
$$
\frac{\partial \Phi}{\partial \xi} =\frac{\partial f(\xi)}{\partial \xi}+\frac{\partial X(\xi)}{\partial \xi}\xi_\mu \xi^\mu, ~~\frac{\partial \Phi}{\partial \xi_{\mu}} =2Xg^{\mu\nu}\xi_{\nu}\, .
$$
For the sake of simplicity, we will not use the arguments of
functions, and the ``prime'' hereafter denotes differentiation
with respect to $\xi$.  To obtain the correct transformation from
the initial model \eqref{act-1}, we must impose the following
constraint,
\begin{equation}
    \det \left( \begin{matrix} \frac{\partial^2 \Phi}{\partial \xi^2} & \frac{\partial^2 \Phi}{\partial\xi \partial\xi_{\mu}} \\
        \frac{\partial^2 \Phi}{\partial \xi_{\nu}\partial\xi} & \frac{\partial \Phi^2}{\partial\xi_{\mu}\xi_{\nu}}\end{matrix}\right) = \det \left( \begin{matrix} f'' +X''\xi_{\mu}\xi^{\mu} & 2X'g^{\mu\nu}\xi_{\nu}\\ 2X'g^{\mu\nu}\xi_{\mu} & 2Xg^{\mu\nu} \end{matrix} \right) \ne 0\, ,
    \label{cond1}
\end{equation}
or equivalently,
$$
2g^{\mu\nu}\left(Xf''+ X'' \xi_\alpha \xi^\alpha \right)-4(X')^2\xi^\mu \xi^\nu \ne 0
$$
In effect, the trace of \eqref{cond1} is,
\begin{equation}
    2Xf'' + \xi_{\alpha}\xi^{\alpha}\left( 2X''X - X'^2 \right) \ne 0\, .
    \label{tracecond1}
\end{equation}
Let us check the constraint for the Starobinsky-Podolsky action
\begin{equation}
    S = \int d^4x \sqrt{-g}\left[R+\frac{c_0}{2}R^2 + \frac{c_1}{2}R_{,\mu}R^{,\mu}\right] +S_{matter}\, .
    \label{spaction}
\end{equation}
The condition \eqref{cond1} in the case at hand is,
\begin{equation}
    \det \left( \begin{matrix} c_0 & 0 \\ 0 & c_1g^{\mu\nu} \end{matrix} \right) \ne 0\, ,
    \label{spcond}
\end{equation}
while the corresponding trace gives,
$$
4c_0c_1 \ne 0\, .
$$
Now we introduce the scalar field $\phi$ and the vector fields $\phi^{\mu}$, defined as follows,
\begin{eqnarray}
    \phi = \frac{\partial\Phi}{\partial \xi} = f' + X'\xi_{\mu}\xi^{\mu} \\
    \phi^{\mu} = \frac{\partial\Phi}{\partial \xi_{\mu}} = 2X\xi^{\mu}\, .
    \label{phiphimu}
\end{eqnarray}
Note that the condition \eqref{cond1} ensures that $ \xi=\xi(\phi,\phi^\mu)$
and $ \xi^\nu=\xi^\nu(\phi,\phi^\mu)$. The scalar potential $U(\phi,\phi_\mu)$  \cite{Cuzinatto:2016ehv} is reduced to,
\begin{equation}
    U(\phi,\phi_{\mu}) = \phi\xi + \phi^{\mu}\xi_{\mu} - \Phi(\xi,\xi_{\mu}) = \phi\xi - f + \frac{\phi^{\mu}\phi_{\mu}}{4X}\, .
    \label{U1}
\end{equation}
The action integral $S' = S'(\phi,\phi^{\mu}, R, R_{,\mu})$ is cast as follows,
\begin{equation}
S' = \int d^4x \sqrt{-g} \left[\phi R + \phi^{\mu}R_{,\mu} - \phi \xi + f - \frac{\phi^{\mu}\phi_{\mu}}{4X}\right] +S_{matter}\, ,
    \label{act3}
\end{equation}
where the scalar and the vector fields $\xi $ and $\xi_\mu$ are
the fundamental fields, and $R$ along with its derivative are
considered as variables. The theory with action \eqref{act3}, is
the scalar-vector-tensor equivalent theory to the theory with the
action \eqref{act-1}. By using the relation
$\sqrt{-g}\phi^{\mu}R_{,\mu} = \partial_{\mu}\left(
\sqrt{-g}\phi^{\mu}R \right) - \sqrt{-g}\nabla_{\mu}\phi^{\mu} R$
and by introducing the new field $\psi = \left( \phi -
\nabla_{\mu}\phi^{\mu} \right)$, after some algebra, the action
\eqref{act3} can be cast as follows,
\begin{equation}
    S' = \int d^4x \left[\sqrt{-g}\psi R - \xi\left( \psi+\nabla_{\mu}\phi^{\mu} \right) + f(\psi,\phi_{\mu},\nabla_{\mu}\phi^{\mu}) -\frac{\phi^{\mu}\phi_{\mu}}{4X}\right] + S_{matter}\, .
    \label{act4}
\end{equation}
Thus the scalar-tensor equivalent theory of \eqref{act-1} is the following,
\begin{equation}
S' = \int d^4x \sqrt{-g}\left[\psi R - U(\psi,\phi_{\mu},\nabla_{\mu}\phi^{\mu})\right] + S_{matter}\, ,
    \label{act5}
\end{equation}
where the scalar potential $U$ is,
\begin{equation}
    U(\psi,\phi_{\mu},\nabla_{\mu}\phi^{\mu}) = \xi\left( \psi+\nabla_{\mu}\phi^{\mu} \right) - f +\frac{\phi^{\mu}\phi_{\mu}}{4X}\, ,
    \label{u3}
\end{equation}
and $\xi$, $f$ are functions of the following arguments,
\begin{eqnarray}
    \xi &=& \xi(\psi,\phi_{\mu},\nabla_{\mu}\phi^{\mu})\\
    f &=& f(\psi,\phi_{\mu},\nabla_{\mu}\phi^{\mu})
    \label{xi}
\end{eqnarray}
It should be noted here that we obtained a Brans-Dicke-like theory
without a kinetic term for the corresponding Brans-Dicke scalar.
By variation of the action \eqref{act5}, one gets the field
equations \cite{Cuzinatto:2016ehv},
\begin{eqnarray}
    \psi G_{\mu\nu} - \left( \nabla_{\mu}\nabla_{\nu}\psi - g_{\mu\nu}\Box\psi \right) + \frac{1}{2}g_{\mu\nu}U = kT_{\mu\nu}\, ,\\
    \label{stFieldEq1}
    R = \frac{\partial U}{\partial \psi}\\
    \frac{\partial U}{\partial \phi^{\rho}}  - \nabla_{\mu}\frac{\partial U}{\partial \left( \nabla_{\mu}\phi^{\rho} \right)} = 0\, ,\\
    \label{stFieldEq2}
    \frac{\delta L_{M}}{\delta \varphi} = 0\, .
\end{eqnarray}
Substituting the potential from Eq. \eqref{u3} into Eq. \eqref{stFieldEq1} we obtain,
\begin{equation}
    \psi G_{\mu\nu} - \left( \nabla_{\mu}\nabla_{\nu}\psi - g_{\mu\nu}\Box\psi \right) + \frac{1}{2}g_{\mu\nu}\left( \xi\left( \psi+\nabla_{\mu}\phi^{\mu} \right) - f + \frac{\phi^{\mu}\phi_{\mu}}{4X} \right) = kT_{\mu\nu}\, .
    \label{stFieldEq11}
\end{equation}
By contracting the above, we obtain the following relation,
\begin{equation}
    -\psi\xi + 3 \Box\psi + 2 \left( \xi\left( \psi+\nabla_{\mu}\phi^{\mu} \right) - f + \frac{\phi^{\mu}\phi_{\mu}}{4X} \right) - kT = 0\, .
    \label{stFieldEq12}
\end{equation}
After some algebra, we finally obtain,
\begin{equation}
\psi\xi + 3\Box \psi + 2 \xi \nabla_{\mu}\phi^{\mu} - 2f + \frac{\phi^{\mu}\phi_{\mu}}{2X} - kT = 0\, .
\label{stFieldEq13}
\end{equation}

For the Starobinsky-Podolsky action (\ref{spaction}), Eq. \eqref{stFieldEq13} transforms to,
\begin{equation}
    -3\Box\psi -kT - 2\left( \frac{1}{2c_0}\left( \psi + \nabla_{\mu}\phi^{\mu} - 1 \right)^2 + \frac{\phi^{\mu}\phi_{\mu}1}{2c_1} \right) + \psi\frac{1}{c_0}\left( \psi + \nabla_{\mu}\phi^{\mu} - 1 \right) = 0\, .
    \label{stFieldEq13SP}
\end{equation}
Now let us derive the equation \eqref{stFieldEq2} using
\eqref{u3},
\begin{multline}
    \frac{\partial \xi}{\partial \phi^{\rho}}\left( \psi + \nabla_{\mu}\phi^{\mu} \right) - \frac{\partial f}{\partial \phi^{\rho}} + \frac{2\phi_{\mu}X - \frac{\partial X}{\partial \phi^\rho}\phi^\mu\phi_\mu}{4X^2} - \nabla_{\mu}\left( \frac{\partial \xi}{\partial (\nabla_{\mu}\phi^\rho)}(\psi + \nabla_{\mu}\phi^\mu) + \xi -\frac{\partial f}{\partial (\nabla_{\mu}\phi^\rho)} - \frac{\phi^\mu\phi_\mu}{4X^2}\frac{\partial X}{\partial (\nabla_{\mu}\phi^\rho)} \right)\, .
    \label{stFieldEq22}
\end{multline}
For the Starobinsky-Podolsky action \eqref{stFieldEq22}, the result is the same as in Ref. \cite{Cuzinatto:2016ehv}, that is,
\begin{equation}
    \frac{\phi_{\mu}}{c_1} - \nabla_{\mu}\left( \frac{1}{c_0}\left( \psi +\nabla_{\nu}\phi^\nu -1 \right) \right) = 0\, .
    \label{stFieldEq22SP}
\end{equation}
An issue we did not address is the occurrence of ghost
instabilities, as we mentioned briefly some sections earlier.
Although ghosts are in general absent, due to the appropriate
choice of the kinetic terms, as was also shown in Ref.
\cite{Naruko:2015zze}, however these instabilities could make
there presence visible via the fluctuations, and particularly in
the sound speed of these fluctuations. It is possible that the
sound speed becomes super-luminal. In addition, one relevant issue
would be to check for gradient instabilities, as for example in
the multi-Galileon theories \cite{Akama:2017jsa}, see also
\cite{Yoshida:2018kwy}. In addition we should note that in the
Einstein frame there are 4 degrees of freedom, to a perturbative
level, due to the presence of a propagating degree of freedom
related to the higher derivatives of the Ricci scalar. The
viability of the whole theoretical framework crucially depends on
this study, so this issue should be carefully addressed in a
focused future work, since it lies beyond the scopes of this
introductory to the subject paper.

\section{Concluding Remarks}

In this paper we investigated the cosmological implications of a
kinetic scalar curvature-corrected $f(R)$ gravity. Particularly,
we included first derivatives of the Ricci scalar in the $f(R)$
gravity action, and we derived the resulting gravitational
equations of motion. We realized various cosmological evolutions
corresponding to a flat FRW Universe, and we demonstrated how the
gravitational equations can be used as a reconstruction technique
for realizing various cosmological scenarios. The interesting
feature of the kinetic Ricci $f(R)$ gravity is that cosmological
evolutions such as bouncing cosmologies and quasi-de Sitter
evolutions, can be realized by a theory which is an
Einstein-Hilbert theory in the $f(R)$ gravity part.

What we did not address in this paper is the study on the
evolutions of cosmological perturbations in the Jordan frame. Such
a study is demanding but compelling in order to find explicit
forms of the spectral index of the primordial curvature
perturbations and of the scalar-to-tensor ratio, and work is in
progress along this research line.

Furthermore, another important issue we need to discuss, is
related to phenomenological aspects of the kinetic Ricci theory,
and particular the question is how to distinguish this theory from
other modified gravity theories. Perhaps the effective field
theory approach may shed some light on this question, as for
example in Refs. \cite{Li:2018ixg,Cai:2018rzd}, where the
effective field theory approach was found to be able to tell the
difference between one particular class of modified gravity
theories from general relativity and other theories. In general,
this study is compelling in many theoretical cosmology contexts,
and along with the effective field theory approach, the
gravitational wave generation of each theory, may also provide
useful hints towards distinguishing each theory. Finally, another
way to distinguish various modified gravities is to study the
growth factor of matter perturbations during the matter domination
era, see for example \cite{Oikonomou:2014gsa}.

Finally, in this paper only inflationary cosmological solutions
were considered and reconstructed. However, a major alternative
theoretical framework able to produce a nearly scale invariant
framework is that of bouncing cosmology
\cite{Cai:2007qw,Cai:2009in,Cai:2012va,Cai:2014bea,Brandenberger:2016vhg}.
The bounce cosmology paradigm has the appealing feature of
producing a cosmological evolution without the unappealing feature
of having an initial singularity. This class of solutions should
also be considered and we hope to address this issue in a future
work.

\section*{Acknowledgments}

This work is supported by MINECO (Spain), FIS2016-76363-P (S.D.O)
and also by project 2017SGR247 (AGAUR, Catalonia) (S.D.O). A.V.N.
thanks the University of KwaZulu-Natal and the National Research
Foundation of South Africa for financial support. Also this
research was supported in part by Russian Ministry of Education
and Science, project No. 3.1386.2017 (S.D.O).

\end{document}